\PassOptionsToPackage{table}{xcolor}
\documentclass[sigconf]{acmart}
\AtBeginDocument{%
  }

\usepackage{microtype}                 % use micro-typography (slightly more compact, better to read)
\usepackage{textcomp}                  % use better special symbols
\usepackage{multirow}   
\usepackage{enumitem}
\usepackage{amsmath}
\copyrightyear{2025}
\acmYear{2025}
\setcopyright{cc}
\setcctype{by}
\acmConference[VRST '25]{31st ACM Symposium on Virtual Reality Software and Technology}{November 12--14, 2025}{Montreal, QC, Canada}
\acmBooktitle{31st ACM Symposium on Virtual Reality Software and Technology (VRST '25), November 12--14, 2025, Montreal, QC, Canada}
\acmDOI{10.1145/3756884.3766001}
\acmISBN{979-8-4007-2118-2/2025/11}
\acmSubmissionID{1245}

\begin{document}

%\title{User performance and behavioral analysis: Hand and tangible interaction in a tabletop Mixed Reality environment}
%\title{User performance and interaction pattern analysis: Hand and tangible interaction in a tabletop Mixed Reality environment}
\title{A Study of Performance and Interaction Patterns in Hand and Tangible Interaction in Tabletop Mixed Reality}

\author{Carlos Mosquera}
\email{carlos.mosquera\_villanueva@siemens.com}
\affiliation{%
  \institution{Siemens A.G.}
  \city{Munich}
  \country{Germany}
}
\affiliation{%
  \institution{Graz University of Technology}
  \city{Graz}
  \country{Austria}
}

\author{Neven Elsayed}
\email{nelsayed@know-center.at}
\affiliation{%
  \institution{Know-Center GmbH}
  \city{Graz}
  \country{Austria}
}

\author{Ernst Kruijff}
\email{ernst.kruijff@h-brs.de}
\affiliation{%
  \institution{Bonn-Rhein-Sieg University of Applied Sciences}
  \city{Sankt Augustin}
  \country{Germany}
}

\author{Joseph Newman}
\email{joseph.newman@siemens.com}
\affiliation{%
  \institution{Siemens A.G.}
  \city{Munich}
  \country{Germany}
}

\author{Eduardo Veas}
\email{eveas@tugraz.at}
\affiliation{%
  \institution{Graz University of Technology}
  \city{Graz}
  \country{Austria}
}

\renewcommand{\shortauthors}{Mosquera et al.}

\begin{abstract}
%What
This paper presents a comprehensive study of virtual 3D object manipulation along 4DoF on real surfaces in mixed reality (MR), using hand-based and tangible interactions. A custom cylindrical tangible proxy leverages affordances of physical knobs and tabletop support for stable input. We evaluate both modalities across isolated tasks (2DoF translation, 1DoF rotation/scaling), semi-combined (3DoF translation+rotation), and full 4DoF compound manipulation.

%Results
We offer analyses of hand interactions, tangible interactions, and their comparison in MR tasks. For hand interactions, compound tasks required repetitive corrections, increasing completion times—yet surprisingly, rotation errors were smaller in compound tasks than in rotation-only tasks. Tangible interactions exhibited significantly larger errors in translation, rotation, and scaling during compound tasks compared to isolated tasks. Crucially, tangible interactions outperformed hand interactions in precision, likely due to tabletop support and constrained 4DoF design. These findings inform designers opting for hand-only interaction (highlighting trade-offs in compound tasks) and those leveraging tangibles (emphasizing precision gains despite compound-task challenges).

\end{abstract}


%\ccsdesc[500]{Hardware~Haptic devices}

%\keywords{Human-centered computing, Mixed Reality, Tangible Interactions, Physical Prop, Haptic devices, Tangible User interfaces}
\begin{teaserfigure}
  \includegraphics[width=\textwidth]{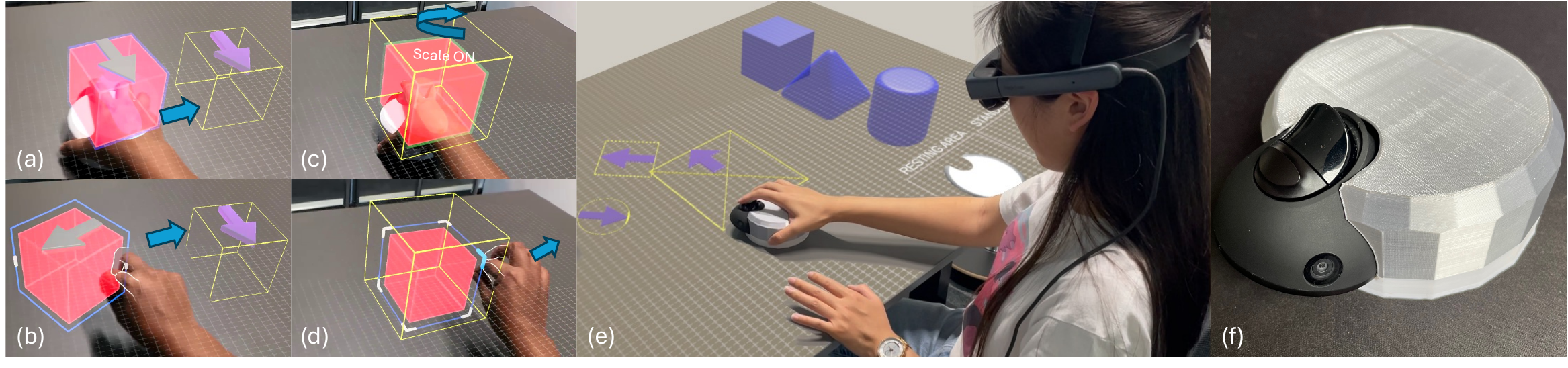}
  \caption{Tasks: (a) Semi-Combined task: Move and rotate using TAN, (b) Semi-Combined task: Move and rotate using HAND, (c)Scaling using TAN, (d) Scaling using HAND, (e) Combined tasks (Move, rotate and Scale) with a puzzle. (f) Final tangible Prop}
  \label{fig:main}
  \Description{Six-part figure illustrating different interaction tasks. (a) and (b) show semi-combined tasks involving movement and rotation using TAN and HAND respectively. (c) and (d) show scaling tasks using TAN and HAND. (e) shows a combined task involving movement, rotation, and scaling with a puzzle. (f) displays the final tangible prop used in the study.}
\end{teaserfigure}

%\received{14 July 2025}

\maketitle

\section{Introduction}

%Context and background:
%-Intro to MR technology
%-Current interaction method(HAND) in MR, and its flaw
Mixed Reality (MR) environments seamlessly blend the physical and virtual worlds, allowing users to interact with virtual objects that can also directly engage with the physical characteristics of the environment \cite{mrdef,mrdeftwo}.  Commercial MR devices, e.g., HoloLens, Magic Leap, Meta Quest, and Apple Vision Pro, have recently been increasingly adopted. These devices predominantly use mid-air hand interaction as the standard input modality, as it is more natural and intuitive. Despite its practicality, this form of interaction lacks tactile feedback, reducing the sense of realism and immersion.

%Intro of TUIs
To address this limitation, researchers have explored Tangible User Interfaces (TUIs), which enhance user interaction by leveraging physical objects to control digital information \cite{brickstan,tantax}. TUIs offer intuitive, direct manipulation capabilities, providing tactile feedback and leveraging physical affordances. These features can improve performance and ease of use in virtual object manipulation tasks \cite{tanisfaster}, although additional external hardware might be needed.

%Intro to Object manipulation
To better understand the potential of TUIs in these contexts, it is essential to consider the fundamental types of 3D object manipulation they are designed to support (iconic tasks): translation, rotation, and scaling \cite{3DObjManSurvey,artaxs}. Selection and deselection are intrinsic requirements for initiating and concluding these manipulations.
%Use case, and why tabletop is relevant
Object manipulation is the core action for many use cases in desktop 3D computing \cite{3DObjManSurvey}. For instance, industrial plant simulation software\cite{plantsimulation} allows engineers to design and optimize production facilities by manipulating 3D assets. When transitioned to MR environments, users can exploit the spatial properties and collaborative opportunities to reach faster design iterations \cite{manuapp}. In this context, tabletops are a good match as they are already normally present in these environments, and they play a crucial role in social gatherings and collaborative activities \cite{tabletopErgo}.   
Therefore, evaluating hand and tangible interactions for core 3D tasks is crucial to ensure they match the affordances and contextual demands of MR tabletop use.
%Therefore, evaluating the appropriateness of Hand and TUIs for these three tasks is of direct interest in MR environments. 

%Research Gap
%Precision not conclusive
Previous studies \cite{remoteobjeval, tanobjeval, smartproxy, inhandball, tansphere} comparing mid-air and TUIs show faster task completion with TUIs but inconclusive results on precision. Despite TUIs often being designed for tabletop use, many evaluations overlooked this context. For example, Kim \cite{inhandball} excluded the tabletop, while Englmeier \cite{tansphere} included it but found no precision gains. Yet, near-distance studies \cite{embodiedaxes} suggest tangibles may offer higher precision. Moreover, scaling, though a core manipulation, remains underexplored. Bozgeyikli \cite{tanobjeval} omitted scaling and complex tasks, also reporting no error differences. 
Approaches~\cite{tansphere, inhandball} addressing scaling rely on prop rotation using affordances of a spherical prop, but lack comparison with alternative methods. These gaps motivate our systematic evaluation of TUIs and mid-air input.

%The way we will investifate
%Research Question
This paper introduces a knob-shaped TUI for precise manipulation enhancing the tabletop experience. We explain the design choices, and compare the tangible with a standard method of interaction in MR applications, namely hand interaction. These interaction types differ fundamentally: the tangible exploits physical properties of the environment (table surface), hand interactions are performed in mid-air. We aim to understand these differences through each individual manipulation task as well as in combination.

%Roadmap
%Contribution
To summarize, our main contributions are: 
(1) The design principles and introduction of a new knob-shaped tangible proxy for 4DoF tasks that leverages the tabletop condition.
(2) Empirical analysis determining the best scaling technique using the proposed tangible proxy, based on performance metrics and user preferences.
(3) Empirical analysis of the effect of task type (Move, Rotate, and Scale) on performance metrics for the tangible proxy and state-of-the-art hand interactions in MR.
(4) Empirical comparison focusing on precision (error metrics) of tangible vs hand interactions.
(5) Analysis of interaction patterns during task execution for both tangible and hand interactions.
%\vspace{-5pt}
\section{Related work}
%previus study
This research builds upon the study of interaction techniques for manipulating virtual objects and tangible user interfaces (TUIs) in tabletop and immersive environments (augmented reality (AR), virtual reality (VR), and mixed reality (MR)). 

\subsection{TUIs on tabletops}  
%Introduction
Tabletops and TUIs support intuitive digital interactions \cite{brickstan}. Interactive tabletops, on which digital content is projected \cite{urp} or where the table itself acts as a reactive screen\cite{reactable}, have showcased the capabilities of TUIs with ongoing research since the mid-1990s \cite{tangiblebits, brickstan}, highlighting their evolving applications and their potential. While utilized in MR environments for some time, studies directly comparing tangibles with conventional MR interaction methods, such as hand gestures and voice commands, remain limited. 

Several innovative TUIs focusing on tabletop interaction techniques have been developed over the years. Tangible Bots \cite{tanbots} used motorized cylinder-shaped tangibles for fine-grained manipulation, especially for rotation. Similarly, Voelker \cite{knobtan} evaluated tangible knobs, finding them to be faster than their virtual counterparts.

%Tabletop AR
Projects like MetaDESK \cite{metadesk} and ARTable \cite{contexttabletan} leveraged TUIs to extend the conventional interactive tabletop to tabletop AR. Tabletop TUIs are not limited to virtual object manipulation. Ssin \cite{geotan} proposed a tangible ring to visualize geo-temporal data on a AR tabletop. Hubenschimid \cite{tanvisar} proposed mobile devices for data visualization on tabletops.
%Challenges and Limitations 
Adapting tabletop TUI techniques to MR presents several challenges, including the need for precise tracking and the integration of physical and virtual elements.

%Subsection object manipulation
\subsection{TUIs for object manipulation in MR}
%Introduction to Early Work: 
Before wearable MR became commercially available, multiple TUI approaches were proposed in the field of wearable AR. Kato \cite{magiccup, magiccupeval} demonstrated a tangible “magic cup” used to grasp virtual objects on a tabletop. Similarly, paddles and wands \cite{padle,tangibleempiricaleval} have been used to manipulate virtual objects. %For example, Billinghurst \cite{tangiblear} showcased a tangible paddle for picking and moving virtual objects placed on a page. 
Potts \cite{tantouch} even proposed a toolkit for object manipulation using gestures on tangible surfaces. Recent research has explored the concept of opportunistic TUIs \cite{oportunistictan2,oportunistictan,smartobj}, appropriating everyday objects for interaction. These tangible approaches are typically evaluated in terms of manipulation tasks.

%Evaluation of Custom Tangibles: 
Among the possible 3D spatial transformations, translation, rotation, and scaling are widely recognized as the primary manipulation tasks \cite{3DUI,3DObjManSurvey}. Translation and rotation, in particular, are generally preferred for evaluating tangible interaction in immersive environments. %TODO: comparition with what? controller hand
Evaluations of custom tangibles for object manipulation in AR/VR environments have been conducted to some extent. Ha \cite{tangibleempiricaleval} empirically evaluated the aforementioned magic cup and paddle for translation tasks . Bowman \cite{remoteobjeval} evaluated various techniques for grasping remote objects in VR. 
%Evaluation of Tangible Proxies: 
For tangible proxies, Reifinger \cite{transroteval} proposed a tracked tangible box with buttons and positively compared it with mouse/keyboard and gesture-based input. Sun \cite{smartproxy} evaluated an image-based tracked tangible cube for grasping objects and compared it to a counterpart with active haptics. Recently, Bozgeyikli \cite{tanobjeval} positively evaluated a tracked tangible cube against controllers and hand interaction for translation and rotation tasks.
%Research on Translation, Rotation, and Scaling: 
Research has also included scaling to some extent. Kim\cite{inhandball} proposed a squeezable ball-shaped controller to perform all three interactions in VR, comparing it with controllers. Kim used a scaling method involving rotating the ball while being squeezed. Englmeier\cite{tansphere} proposed a buttonless tangible sphere for AR, where rotation was also used for scaling. Englmeier compared all interaction tasks with controllers, under both table and no-table conditions.

These studies have consistently shown that tangibles improve completion time compared to mid-air counterparts. However, detailed analysis of how individual tasks affect metrics is still lacking. Furthermore, no evidence has been found for precision improvement. Therefore, we propose this study to address these gaps.

\section{Study Design}
Manipulating 3D objects is a key aspect of MR, where virtual elements must seamlessly interact with the real world. A common use case relying heavily on 3D object manipulation is constrained world-at-scale modeling, e.g. applications like virtual furniture placement \cite{room1,room2}, floor layout design, and industrial plant planning \cite{plantsimulation}. We look at the latter as a guiding use case without loss of generality. 

\subsection{4DoF Use case}
%Description
Within the scope of our requirements engineering and design process, we considered industrial plant planning (plant layout design) as a reference use case. It involves selection and manipulation tasks at varying performance granularities, common across domains.
Engineers use desktop 3D software to design plant layouts by placing machines on a virtual shop floor, optimizing workflows and personnel movement. This helps to identify bottlenecks early, shortens design cycles, and creates a digital twin of the physical plant. 

Despite their extensive features, these tools remain complex for non-experts. Plant design involves stakeholders with diverse technical expertise, yet current tools lack intuitive interfaces and collaborative support. To explore the use case, a handheld AR prototype using tabletop tangibles was built. A scaled plant layout is placed on a table, with machines manipulated via a QR-tracked prop—enabling natural interaction and visualization for non-experts.

One-on-one interviews were conducted with three industrial planning experts. They positively reviewed the prototype and provided key requirements for object manipulation. These informed both the tangible design and the study setup. Machines, typically resting on surfaces and subject to gravity, should be operable intuitively and precisely—favoring 2DoF movement along the surface. Objects are usually upright, naturally limiting rotation to 1DoF around the vertical axis. And, since objects maintain a fixed aspect ratio, uniform scaling is preferred (1DoF). While more complex manipulations are possible, these 4DoF represent most interactions. These insights guided the tangible prop design, aiming for precise and intuitive manipulation for the identified 4DoF constraints.

\subsection{Tangible prop design}

%Why a cylinder shape
Multiple considerations were taken into account in the design of the tangible prop. Prior studies \cite{sizeshapear,sizeforar,sizeforar2} demonstrate that having matching shapes and scales for objects are crucial factors in the design of tangible proxies. Instead, our goal was to create a general-purpose tangible that can manipulate objects of various shapes. 
Rau \cite{generaltan} highlighted handling and affordance as the main design requirements for tangibles in mobile AR. Following this principle, spherical-shaped TUIs were successfully tested \cite{tansphere, inhandball}. Spherical TUIs have a uniform shape without a clear upright position. They can be rotated along all axes simultaneously (3DoF rotation), but it might prove difficult to exert precise control over each axis separately. Spherical tangibles are conveniently handheld but lack stability on surfaces. We opted for a flat-bottomed shape to take advantage of the support the tabletop offers. 

A cylinder shape retains the rounded affordance of a sphere while leveraging the characteristics of knobs for fine-grained input of magnitudes. Therefore, we explore a hand-sized cylindrical shape, which can be easily grasped and lifted with one hand, inviting the user to place it on the table and use the curved lateral surface to rotate it, similar to a knob \cite{knobtan}.

%Optional
%Tracking 
Tracking the prop was challenging, as reliable tracking is essential for performance and user experience \cite{desigintmobilear}. We tested several methods: Vuforia \cite{generaltan} was unreliable due to hand occlusion; VR trackers \cite{vrtrackertoar} complicated alignment and reduced ergonomics without improving responsiveness; and Vicon \cite{sizeforar} was too complex. Ultimately, we embedded the Magic Leap 2 controller in the prop \cite{tanobjeval}, which provided adequate tracking with minimal integration.

%Input 
Another key aspect was the "selection" input—detecting user intent to start manipulation. To keep interaction simple and focused, we used a physical button with toggle behavior: one press to engage, another to disengage \cite{select}. This was especially important for the scale-by-lifting method, where holding a button while lifting was hard.
For scaling, we mapped direct prop actions (e.g., moving or rotating) to scaling, requiring a second input to switch modes. The trackable controller, conveniently oriented, was ideal—offering two buttons for both selection and mode switching in a compact, intuitive form.

%3D printed
The final tangible prop (Figure~\ref{fig:main}f) was 3D printed in two parts for easy assembly, with an internal cavity to house the controller—used for both tracking and input. The design kept the tracking cameras unobstructed and buttons accessible. Minor refinements included trimming the top edges for comfort and adding a cloth base for smoother sliding. The assembled prop weighs 270g and measures 11cm in diameter and 4.7cm in height.

\subsection{Apparatus}
The Magic Leap 2 MR headset \cite{magicleap2} was used, consisting of a computer pack (AMD Zen 2 CPU, 16GB RAM, AMD RDNA 2 GPU) connected via cable to an optical see-through display (FOV: 75°H ×70°V, 544×470px). It includes a 6-DoF controller (vision-based and IMU tracked)\cite{magicleap2controller}.
The tangible prop was custom 3D printed using PLA on an Ultimaker printer. A USB foot pedal was also used.

\subsection{Interaction Design}
This section outlines the interaction design used in the main study, focusing on the two input modalities: hand-based interaction (HAND) and tangible-based interaction (TAN). 

\subsubsection{HAND Interactor} 
To evaluate hand interactions for object manipulation, we selected a state-of-the-art library using the Mixed Reality Toolkit 3 (MRTK3)\cite{MRTK3} for Magic Leap 2, choosing the near interaction \mbox{\textit{"BoundsControl"}} implementation\cite{MRTK3Boundbox}.

\textit{Move and Rotate by Pinch gesture:} 
The interaction involved near interaction with the virtual object and the handles positioned around it. To move the object, the user pinches inside it and moves their hand. Rotation can be performed simultaneously by rotating the hand or by pinching and dragging the sides handle. Releasing the pinch completes the action. (see Figure\ref{fig:main}b)

\textit{Scaling by Pinch gesture:} 
To scale the object, the user pinches a corner handle (see Figure\ref{fig:main}d) and moves their hand while maintaining the gesture. Releasing the pinch finalizes the scaling.

%\vspace{-5pt}
\subsubsection{TAN Interactor} 
The first point to consider is object selection. A physical button -- \textit{the interaction button}-- acts as a simple engaging button\cite{select} which the user presses once to start interacting with an object and presses it again to disengage. A second button was used to control the interaction modality--- \textit{the modal button}.

The cylinder shape resting on the table offers 2DoF motion on its surface and 1DoF knob like rotation. To  support 4DoF object manipulation, it is necessary to account for one missing DoF. 

\textit{Move and Rotate using the Prop:}
For Move and Rotate, the mapping of manipulation can be directly mapped to the natural 3DoF: Move is translation in 2DoF on tabletop; rotate around vertical axis is 1DoF rotation of the prop.
The user must place the tangible prop inside the target virtual object. Visual color feedback on the object indicates that object can be selected. To do so, the user presses the interaction button once. Translating or rotating the prop will correspondingly move or rotate the virtual object. The task is finished by pressing the button again. Translate and rotate happen in parallel.

\textit{Scaling challenge:}
Among the primary types of 3D object manipulation (translation, rotation, scaling), scaling has no direct mapping\cite{3DObjManSurvey}. This presents a challenge when determining an appropriate scaling method using TUIs.
Consequently, we decided to conduct a pilot study to systematically evaluate and identify the most appropriate scaling method for our final experiment.

%\vspace{-5pt}
\section{Pilot Study: Optimal Scaling Method}
Few studies exploring tangible proxies in MR include scaling. Notably, spherical-shaped tangibles have been evaluated for scaling in both VR~\cite{inhandball} and AR~\cite{tansphere}, leveraging the affordance of their form through prop rotation—though they did not formally compare with alternative methods. This pilot compares three tangible scaling methods and a baseline hand scaling. It also serves as a preliminary test of the tangible design and assesses whether interaction performance is consistent across different positions on the table.

\subsection{Pilot Study Design}
To explore intuitive 1DoF scaling, we focused on interactions that map a single directional input to a single scaling dimension. We selected LIFT (vertical motion), MOVE (horizontal translation), and ROT (rotational input)  for their clear spatial or directional mappings, which support precise magnitude control. Each technique engages different arm biomechanics. Four techniques were evaluated:

\textit{HAND \textit{(mid-air)}:} A baseline, where users pinch corner handles of a bounding cube to scale the object.
\textit{LIFT \textit{(mid-air tangible)}:} Users lift the prop inside the virtual object then press a button to activate scaling. Vertical motion increases or decreases size. LIFT engages the shoulder biomechanics.
\textit{ROT \textit{(tabletop tangible)}:} Users rotate the prop clockwise/counterclockwise to scale up/down. ROT engages the wrist biomechanics.
\textit{MOVE \textit{(tabletop tangible)}:} Users translate the prop left/right, mimicking a slider and mapping linear motion to scaling. MOVE engages the elbow biomechanics.

The task involved scaling a red cube to fit a yellow target frame. Two conditions were tested: a scaling direction condition (up or down), with a fixed initial size and varying target sizes; and a starting position condition, where the cube was randomly placed along a semicircular arc (40°, radius = 0.8m) in front of the user—categorized as left, center, or right—to ensure consistent arm reach across trials. %(see Figure \ref{fig:setup}).

The interaction techniques are compared based on performance, assessed through quantitative measures: completion time, error, and overshoot. The effects of scaling direction and object position on performance are analyzed independently for each technique.

\subsection{Pilot Study Results}
19 participants (9 women and 10 men) who reported having no physical or mental condition were voluntarily recruited from the company, age \textit{(M: 33.84 , SD: 10.82)}. 
They rated their experience with AR/VR on a scale of 1 to 10 \textit{(M: 6.0, SD: 2.47)} and their familiarity with the MR glasses \textit{(M: 1.47, SD: 1.07)}. 4032 trials were collected from 18 participants (excluding one left handed participant). 

%A Shapiro-Wilk test indicated non-normality in completion time and error data. Therefore, we used a Kruskal-Wallis test followed by post-hoc Dunn tests for pairwise comparisons. Results are summarized in Figure~\ref{fig:taskperformancescale}, with further details in the supplementary material.
A Shapiro-Wilk test indicated non-normality in completion time and error data. Therefore, we used a Friedman test, followed by post-hoc Wilcoxon tests for pairwise comparisons. Results are summarized in Figure~\ref{fig:taskperformancescale}, with details in the supplementary material.

The findings show that ROT: Scaling by rotation led to superior task performance, supported by objective metrics (completion time and error rates) and user preference. The position condition had no significant effect and was excluded from the final study design.

Overall, the pilot study validated the TUI design, confirmed user acceptance, and informed the interaction design for the main study.

\textit{Scale using the Prop:} The final scale interaction in the main study unfolds as follows: selecting the virtual object by placing the prop inside it while it rest on the table, and click the selection button; the modal button switches between "scaling" and "rotating". In scaling mode, rotating the prop will scale the object (see Figure \ref{fig:main}c).

%overshoot counts,
\begin{figure}
    \centering
    \includegraphics[width=0.9\linewidth]{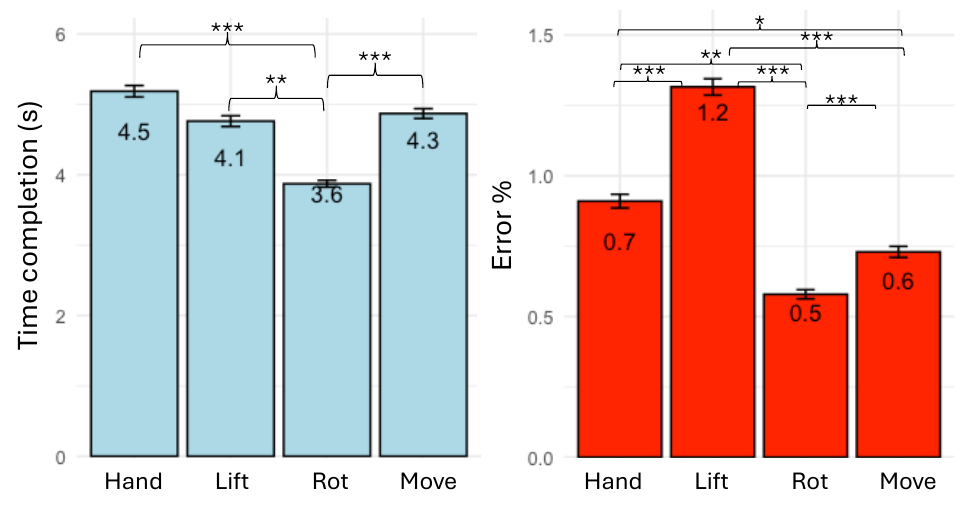}
    \caption{Scaling Pilot Results: Effect of interaction technique on performance. Left: Completion Time, Right: Error}
    \Description{Bar charts showing the effect of interaction technique on performance in a scaling pilot study. The left chart displays task completion time, and the right chart shows error rates for different techniques.}
    \label{fig:taskperformancescale}
\end{figure}

\section{ Main Study: Full Manipulation Pipeline}

\subsection{Task Design}
The main study comprises three main tasks for manipulating a virtual object: Move and Rotate, Only Scale, and Combined. These tasks are split into five operations: three isolated tasks (Only Move, Only Rotate, Only Scale) and two mixed tasks (Semi-Combined: Move and Rotate, Combined: Move, Rotate, and Scale).

\textit{TASK 1: Move and Rotate.}
For this task, the subject was required to move and rotate a red cube to match a goal box represented by a yellow frame. Both the red cube and the goal box have an arrow on top to indicate orientation, which the subject is also required to match. The move task was 2D(x, y), while the rotation was 1D around Z-axis. Within this task, sub-tasks depend on the interaction required : T1.1 Only-Move: The red cube and goal frame appeared in different positions, but their arrows aligned, so no rotation was required. T1.2 Only-Rot, the red cube and goal frame were in the same position, the rotation was different. T1.3 Move-And-Rotate, combines both. Figure \ref{fig:main}a and \ref{fig:main}b show this task in action.

\textit{TASK 2: Only Scale.}
Participants were asked to scale the red cube to match the size of the goal box (a yellow frame). The task is a 1D uniform scale (in all axis). The tasks contained scaling up or down the red cube.
Figure \ref{fig:main}c and \ref{fig:main}d illustrates the task.

\textit{TASK 3: Combined.}
The last task was a combination of both Task 1 and Task 2, but in a puzzle format to add more cognitive complexity. Participants were presented with a pool of 3 shapes (Cube, Cylinder, Pyramid) located on the right side of the table and many goal blueprints distributed on the table. The blueprints contained the required shape and the orientation. The participant had to first spawn a desired object from the pool by selecting the shape and then manipulate it to match the position, rotation and scale of the respective goal blueprint. Two levels of difficulty of the task were implemented: (1) medium difficult with 3 goal objects, (2) hard with 5 goal objects. Each task was considered as 3 or 5 trials depending on the difficulty. Figure \ref{fig:main}e illustrates this task.

\subsection{Hypothesis}

%ToDo: improve how do we get to the Hypothesis
Cheng \cite{tabletopErgo} demonstrated that tabletop interactions enhance ergonomics and comfort compared to mid-air interactions in VR context. Our Empirical observations of the hand and arm biomechanics in the designed interactions reveal that mid-air interactions engage more arm and hand muscles compared to tabletop interactions. This should affect not only the performance but how precise the subject perform, as well as the strategy for more complex taks. Moreover \cite{embodiedaxes} demonstrated how TUI can improve precision compared to mid-air interactions.
Consequently, we formulated the following hypotheses: 

\textit{\textbf{H1}}: Across all tasks, TAN will have better performance compared to HAND, especially for precision. 

\textit{\textbf{H2}}: The interaction type will affect performance per task, as well as user interaction patterns.

\subsection{Study Procedure}
Participants were invited into an isolated room to avoid disturbances. Inside, an adjustable table and chair were provided; subjects performed the experiment seated. The study began with the supervisor welcoming the participant and explaining the purpose. The participant then read and signed a consent form, followed by a pre-questionnaire for demographic and experience data. The supervisor adjusted the chair and table height for comfort. The participant placed a paper marker in the center of the table within easy reach. Next, the supervisor set up the MR glasses, positioned a virtual screen, calibrated interaction areas based on the marker, and set up the tangible prop and foot pedal. The participant was introduced to the MR glasses, which were adjusted to fit. The system explained all study components, with the supervisor available to clarify questions. A training phase allowed participants to familiarize themselves with the interaction methods before starting the main study, which consisted of 160 trials and took about 40 minutes. Participants could move between trials using the foot pedal. Upon completion, the system saved the data and notified the participant. The supervisor then retrieved the MR device, and the participant completed a post-experiment questionnaire. A 7-point NASA-TLX questionnaire for task load was recorded, along with a 7-point Likert scale for preferences regarding each interaction method and overall preferences for tangible and hand-based interactions. Open-ended questions captured qualitative feedback, including suggestions for improvement.
Procedures were approved by the TU Graz Ethics Committee (GZ: EK-41/2024).

\subsection{Measures}
For each trial, different timestamps were recorded: \textit{Visible\_time}, \textit{Select\_time}, \textit{Deselection\_time}, etc. \textit{Completion Time} was calculated: $\text{Last\_deselect\_time} - \text{First\_select\_time}$. \textit{Overshoot} count was calculated by counting how many \textit{Select\_time} were recorded.

Moreover, the task defined the goal conditions: \textit{Goal\_position (vector2)}, \textit{Goal\_rotation}, \textit{Goal\_scale}, represented by a yellow frame within which the red cube had to be fit. The pose of the red cube at task end was recorded: \textit{End\_position (vector2)}, \textit{End\_rotation}, \textit{End\_scale}. From this information, the absolute errors were calculated: \\
%$ \text{Position\_Error} = \text{End\_position} - \text{Goal\_position} $, $ \text{Rotation\_Error} = \text{End\_rotation} - \text{Goal\_rotation} $ and $ \text{Scale\_Error} = \text{End\_scale} - \text{Goal\_scale} $.
$\text{Position\_Error}= \text{End\_position} - \text{Goal\_position};\text{Rotation\_Error} = \text{End\_rotation} \allowbreak\ - \text{Goal\_rotation}; \text{Scale\_Error} = \text{End\_scale} - \text{Goal\_scale}.$

To analyze subject interaction patterns towards the task, transforms of TAN, HAND, and the object being interacted with were recorded every 100 ms as time-series data, also if TAN or HAND were interacting. From the object-being-interacted transform, the active task (Move, Rotate, Scale) being executed is calculated.

\section{Results}
\begin{figure*}
    \centering
%    \includegraphics[width=1\linewidth]{pictures/All2.pdf}
%    \caption{Results for HAND and TAN: (Left) Pair-wise Dunn test results for metrics across tasks, (Middle) Sequential Analysis for Combined task from a representative user, (Right) Spatial pattern of interactors from a representative user in the MoveAndRotate task}
    \includegraphics[width=1\linewidth]{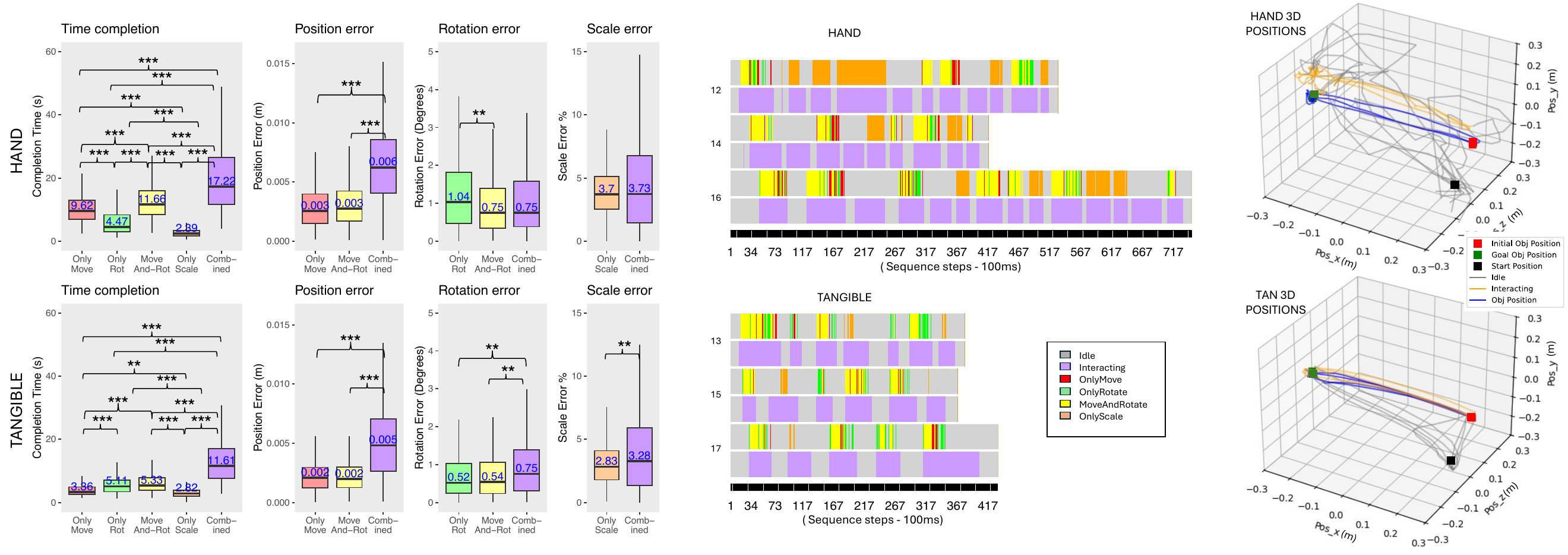}
    \caption{Results for HAND and TAN: (Left) Pairwise test results for metrics across tasks, (Middle) Sequential Analysis for Combined task from a representative user, (Right) Spatial pattern of interactors from a representative user in Only-Move}
    \label{fig:allresults}
    \Description{Three-part figure showing results for HAND and TAN interaction methods. Left: pairwise Wilcoxon test results across tasks. Middle: sequential interaction analysis for the Combined task from a representative user. Right: spatial interaction pattern in the MoveAndRotate task from a representative user.}
\end{figure*}

30 participants (15 women and 15 men) were voluntarily recruited from the university and the company, age \textit{(M: 32.00 , SD: 7.43)}. All of them were graduated and right-handed, 12 wore corrective glasses. They rated their AR/VR experience \textit{(M: 5.73, SD: 2.62)} and their familiarity with the MR glasses \textit{(M: 2.17, SD: 2.53)} on a 1–10 scale, where 1 indicated low and 10 high experience/familiarity. One participant was excluded due to a headache caused by the device, which he had anticipated due to prior VR experience. Participants reported having no physical or mental condition. 

%TODO: Maybe add the ethics document
\subsection{Data Analysis Approach}
We analyzed Hand and Tangible Interaction in terms of quantitative measures and interaction patterns. First, we examined each interaction type separately, followed by a comparative analysis.

%\subsubsection{Quantitative Measures:} We assessed completion time across all tasks. Position error was analyzed for tasks involving position adjustment (Only-Move, Move-and-Rot, Combined), rotation adjustment (Only-Rot, Move-and-Rot, Combined), and scale adjustment (Only-Scale, Combined). A Shapiro-Wilk test indicated non-normal distributions for completion time and errors, so we used the Kruskal-Wallis test with post-hoc Dunn tests for pairwise comparisons when needed. Effect sizes were also computed. For NASA-TLX and preference ratings, we applied the Friedman test followed by Kendall’s W. All analyses were conducted in R.
\subsubsection{Quantitative Measures:} We assessed completion time across all tasks. Position error was analyzed for tasks involving position adjustment (Only-Move, Move-and-Rot, Combined), rotation adjustment (Only-Rot, Move-and-Rot, Combined), and scale adjustment (Only-Scale, Combined). A Shapiro-Wilk test indicated non-normal distributions for completion time and errors, so we used the Friedman test with post-hoc Wilcoxon tests for pairwise comparisons when needed. Effect sizes were also computed. For NASA-TLX and preference ratings, we applied the Friedman test followed by Kendall’s W. All analyses were conducted in R.

\subsubsection{Interaction pattern analysis:} We visually examined a representative sample of a user. We verified that the observed patterns were consistently present across user samples. The analysis focused on two aspects: spatial interaction patterns during the Move-and-Rotate task, and interaction sequences in the Combined task.

\textit{Sequence analysis:} We analyzed how subjects interacted during the complex task (Combined) using the TramineR R package. A sequence of labeled (idle, move, rot, scale) events is extracted from timestamped interaction data and visualized (Figure ~\ref{fig:allresults} center). The Combined task illustrates interaction with 3 objects in a sequence. 

\textit{Spatial analysis:} We plotted the HAND and TAN trajectory for the Only-Move task--four trials with identical start and goal conditions (Figure~\ref{fig:allresults}, right). The plots show the resting position (black box), initial object position (red box), and goal position (green box). Interaction traces show idle (grey) and object motion (blue lines).

\subsection{Hand Interaction}

\subsubsection{Quantitative Measures}
%For HAND, the Kruskal-Wallis test revealed a significant difference in completion time across tasks \textit{($\chi^2(2) = 11148.48, p < 0.001, \eta^2 = 0.56$)}, 
%The follow-up pair-wise Dunn test (Figure ~\ref{fig:allresults} left) revealed that time completion of compound tasks is significantly longer than their individual components, e.g., Move-And-Rot takes significantly longer than Only-Move and that Only-Rot. Across the individual tasks, Only-Move took longer than Only-Rot and Only-Scale, all differences significative. 
For HAND, the Friedman test revealed a significant difference in completion time across tasks \textit{($\chi^2 4= 105.17, p< 0.001, W= 0.94$)}, 
The follow-up pairwise Wilcoxon test (Figure ~\ref{fig:allresults} left) revealed that completion time of compound tasks is significantly longer than their individual components, e.g., Move-And-Rot takes significantly longer than Only-Move and that Only-Rot. Across individual tasks, Only-Move took longer than Only-Rot and Only-Scale, all differences significative.

%The Kruskal-Wallis test revealed a significant difference in position error \textit{($\chi^2(2) = 317.45, p < 0.001, \eta^2 = 0.16$)} when comparing Only-Move, Move-And-Rot, and Combined. Pair-wise Dunn tests revealed that participants incurred significantly larger position errors in Combined compared to Move-And-Rot and also compared to Only-Move. Notably, the difference between Move-And-Rot and Only-Move in position errors was not significant.
The Friedman test revealed a significant difference in position error \textit{($\chi^2 2= 43.14, p< 0.001, W= 0.77$)} when comparing Only-Move, Move-And-Rot, and Combined. Wilcoxon tests revealed that participants incurred significantly larger position errors in Combined compared to Move-And-Rot and also compared to Only-Move. Notably, the difference between Move-And-Rot and Only-Move in position errors was not significant.

%The Kruskal-Wallis test revealed a significant difference in rotation error \textit{($\chi^2(2) = 20.95, p < 0.001, \eta^2 = 0.01$)} when comparing Only-Rot, Move-And-Rot and Combined. Surprisingly, rotation errors were largest (significantly) in Only-Rot, while both compound tasks were comparable in Rot Error.
The Friedman test revealed a significant difference in rotation error \textit{($\chi^2 2= 11.21, p= 0.003, W= 0.2$)} when comparing Only-Rot and Move-And-Rot. Surprisingly, no significant differences were found between Only-Rot and Combined.
There was no significant difference in scaling errors between Only-Scale and Combined. 

\subsubsection{Interaction pattern Analysis}

The \textit{Sequence analysis} reveals a comparatively large number of scale adjustments vs moving and rotating. Rotation adjustments appear more sporadic and short. 

The \textit{Spatial analysis} shows the HAND unrestricted movement is indicative and visible when approaching the object (grey traces arriving in the red box) at diverse angles. The object motion trace (blue) is relatively stable in comparison, but there is still an amount of motion in all 3DoF. Leaving the target (gray traces leaving green box) is again unstable.

\subsection{Tangible Interaction}

\subsubsection{Quantitative Measures}
%A Kruskal-Wallis test revealed significant differences in completion time across tasks  \textit{($\chi^2(2) = 899.72$, $p < 0.001$, $\eta^2 = 0.44$)}. Post-hoc Dunn pair-wise comparisons revealed that Combined takes significantly longer compared to all other tasks. Move-And-Rot takes significantly longer than Only-Move, but is not longer than Only-Rot. 
A Friedman test revealed significant differences in completion time across tasks  \textit{($\chi^2 4= 101.0, p< 0.001, W= 0.9$)}. Post-hoc Wilcoxon pairwise comparisons revealed that Combined takes significantly longer compared to all other tasks. Move-And-Rot takes significantly longer than Only-Move, but is not longer than Only-Rot.

%A Kruskal-Wallis test revealed significant differences in position error \textit{($\chi^2(2) = 220.93.45, p < 0.001, \eta^2 = 0.11$)}. Post-hoc Dunn pair-wise comparisons revealed that position errors were significantly larger in the Combined tasks compared with Move-And-Rot and with Only-Move. No significant difference was observed between Move-And-Rot and Only-Mov in position error.
A Friedman test revealed significant differences in position error \textit{($\chi^2 2= 42.28, p< 0.001, W= 0.75$)}. Post-hoc Wilcoxon pairwise comparisons revealed that position errors were significantly larger in the Combined tasks compared with Move-And-Rot and with Only-Move. No significant difference was observed between Move-And-Rot and Only-Mov in position error.

%Kruskal-Wallis test revealed significant differences in rotation error \textit{($\chi^2(2) = 20.22, p < 0.001, \eta^2 = 0.01$)}. Post-hoc Dunn tests revealed again a significative error in Combined compared with Move-And-Rot as well as with Only-Rot. No significant difference was observed between Move-And-Rot and Only-Rot in rotation error.
Friedman test revealed significant differences in rotation error \textit{($\chi^2 2= 12.07, p= 0.002, W= 0.21$)}. Post-hoc Wilcoxon tests revealed again a significative error in Combined compared with Move-And-Rot as well as with Only-Rot. No significant difference was observed between Move-And-Rot and Only-Rot in rotation error.

%A Dunn test revealed a significant difference in scaling error between Combined and Only-Scale ( see Figure~\ref{fig:allresults}-Left Bottom).
A Wilcoxon test revealed a significant difference in scaling error between Combined and Only-Scale (see Figure~\ref{fig:allresults}-Left Bottom).

\subsubsection{Interaction pattern Analysis}

The \textit{Sequence analysis} reveals short interactions initiated with move-and-rotate actions and followed by rotation adjustments and scaling adjustments. Scaling adjustments are sporadic and short.

The \textit{Spatial analysis} shows that the Tangible resting on the table results in a stable interaction trace. Clear consistent traces are observable at all stages. When approaching the object, gray traces arrive at the red box from a consistent angle. Similar pattern is observed in the translation and release of the object.

\subsection{Comparing Hand vs Tangible}
Table ~\ref{table:tanvshand} details the comparison HAN vs TAN across all tasks and measures. For this comparison, we also measure overshoot count and the subjective feedback about workload with NASA TLX.

\subsubsection{Quantitative Measures}
%Comparing \emph{time-completion}, TAN interaction resulted in significantly shorter completion times in tasks that involve moving the object (Only-Move, Move-And-Rot, Combined) with large and medium effect size, while Only-Scale was faster with HAND, small effect size. Position adjustment errors were consistently significantly lower with TAN than with HAND.
Comparing \emph{completion-time}, TAN interaction resulted in significantly shorter completion times in tasks involving Prop movement (Only-Move, Move-And-Rot, Combined) with large effect size, while Only-Scale was faster with HAND, small effect size. Position adjustment errors were consistently significantly lower with TAN than with HAND.
Rotation adjustment errors were significantly lower with TAN than with HAND in Only-Rot and Move-And-Rot tasks. However, the combined task evidenced an increase in rotation adjustment errors for the TAN, resulting in no significant difference. 
Scale adjustment errors were significantly lower with TAN than HAND in Only-Scale, but no significant differences were found in the Combined task.

\subsubsection{Interaction pattern Analysis}
Comparing HAND and TAN interaction patterns reveals notable differences. In the sequence analysis (with evenly spaced time steps), the shortest interaction sequence for HAND is 417 steps and for TAN is 367. The longest interaction sequence for HAND is 800 steps vs 420 for TAN, almost double the length. The sequence of actions also reveal that participants engage numerous scaling actions in HAND compared with TAN where these are sporadic and short. 
Besides the scaling actions, participants with HAND interaction spend longer in the Move-And-Rotate action. Analyzing the spatial pattern of a Move-And-Rotate task comprising multiple trials with the same origin and destination for the object, we find that TAN has a more distinctive stable pattern supported by the prop resting on the table surface. Whereas HAND is more unstable with motion also spanning vertical axis. In practice, Move actions took longest with HAND.

\begin{table*}[h!]
\centering
\caption{Results of the Friedman Test and Effect size comparing Tangible (TAN) and Hand (HAND) Interaction.}
\small
\begin{tabular}{|l|l|l|l|l|l|l|l|l|l|}
\hline
\textbf{Metric} & \textbf{Task} & \textbf{TAN Median} & \textbf{TAN SD} & \textbf{HAND Median} & \textbf{HAND SD} & \textbf{$\chi^2$} & \textbf{p-value} & \textbf{r} & \textbf{Magnitude} \\
\hline
\multirow{5}{*}{Completion Time(s)} & OnlyMove & 3.31 & 1.42 & 9.78 & 3.11 & 28.00 & \cellcolor{lightgray}{<0.001} & 1.00 & large \\
\cline{2-10}
 & OnlyRot & 5.15 & 2.48 & 5.07 & 3.37 & 0.57 & 0.450 & 0.02 & small \\
\cline{2-10}
 & MoveAndRot & 5.47 & 2.26 & 11.85 & 4.18 & 28.00 & \cellcolor{lightgray}{<0.001} & 1.00 & large \\
\cline{2-10}
 & OnlyScale & 2.83 & 1.09 & 2.31 & 0.86 & 7.00 & \cellcolor{lightgray}{0.008} & 0.25 & small \\
\cline{2-10}
 & Combined & 12.42 & 5.08 & 17.72 & 7.91 & 20.57 & \cellcolor{lightgray}{<0.001} & 0.73 & large \\
\hline
\multirow{3}{*}{Position Error (mm)} & OnlyMove & 1.95 & 0.67 & 2.64 & 0.86 & 5.14 & \cellcolor{lightgray}{0.023} & 0.18 & small \\
\cline{2-10}
 & MoveAndRot & 2.02 & 0.68 & 3.01 & 1.00 & 14.29 & \cellcolor{lightgray}{<0.001} & 0.51 & large \\
\cline{2-10}
 & Combined & 4.84 & 1.30 & 6.37 & 1.53 & 17.29 & \cellcolor{lightgray}{<0.001} & 0.62 & large \\
\hline
\multirow{3}{*}{Rotation Error (Degree)} & OnlyRot & 0.54 & 0.33 & 1.28 & 0.53 & 14.29 & \cellcolor{lightgray}{<0.001} & 0.51 & large \\
\cline{2-10}
 & MoveAndRot & 0.56 & 0.27 & 0.87 & 0.35 & 9.14 & \cellcolor{lightgray}{0.002} & 0.33 & moderate \\
\cline{2-10}
 & Combined & 0.90 & 0.34 & 0.83 & 0.52 & 0.14 & 0.705 & 0.01 & small \\
\hline
\multirow{2}{*}{Scale Error (mm)} & OnlyScale & 2.95 & 0.69 & 3.63 & 0.75 & 11.57 & \cellcolor{lightgray}{<0.001} & 0.41 & moderate \\
\cline{2-10}
 & Combined & 3.57 & 1.03 & 3.81 & 1.75 & 2.29 & 0.131 & 0.08 & small \\
\hline
\end{tabular}
\label{table:tanvshand}
\end{table*}

\subsubsection{Overshoot Count}
%For the Move and Rotate task, A Kruskal-Wallis test indicated that there was a significant difference in overshoot count across interaction \textit{($\chi^2(2) = 420.72, p < 0.001, \eta^2 = 0.49$)}. Here TAN \textit{(M: 1.6, SD: 1.08)} presented less overshoot than HAND \textit{(M: 3.95, SD: 1.73)}. 
For the Move-And-Rotate task, A Friedman test indicated that there was a significant difference in overshoot count across interaction \textit{($\chi^2 1= 19.59, p< 0.001, W= 0.7$)}. Here TAN \textit{(M: 1.57, SD: 0.84)} presented less overshoot than HAND \textit{(M: 3.62, SD: 0.98)}. 
%For the Combined task, A Kruskal-Wallis test indicated that there was a significant difference in overshoot count across interaction \textit{($\chi^2(2) = 353.77, p < 0.001, \eta^2 = 0.30$)}. Here TAN \textit{(M: 3.11, SD: 1.97)} presented less overshoot than HAND \textit{(M: 6.52, SD: 3.99)}.
For the Combined task, A Friedman test indicated that there was a significant difference in overshoot count across interaction \textit{($\chi^2 1= 26, p <0.001, W= 0.93$)}. Here TAN \textit{(M: 2.93, SD: 1.35)} presented less overshoot than HAND \textit{(M: 5.87, SD: 1.48)}.

\subsubsection{NASA TLX \& Preference}
\begin{figure}
    \centering
    \includegraphics[width=1.1\linewidth]{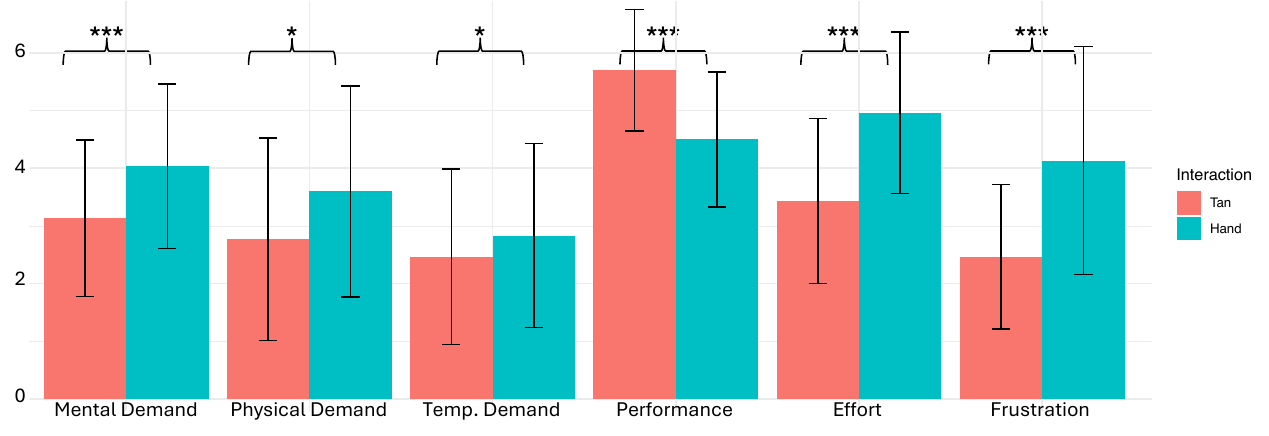}
    \caption{NASA-TLX for the combined task}
    \label{fig:nasa-tlx}
    \Description{Bar chart showing NASA-TLX workload scores for the combined task condition.}
\end{figure}

For task load, Friedman test showed that, in the combined task, mental demand \textit{($\chi^2 2=9,\ p=0.003,\ \text{Kendall} \allowbreak\ \text{-W}=0.3$)}, and performance, effort, and frustration \textit{($\chi^2_2=20.17,\ p<0.001,\ \text{Kendall-W}=0.672$)} were affected by interaction type. Further analysis for other tasks is shown in Figure~\ref{fig:nasa-tlx}.
Friedman tests indicated that Tangible was preferred for the Move-and-Rotate task \textit{($\chi^2 2= 12.46, p= 0.004, \text{Kendall-W}= 0.38$)} and Combined task \textit{($\chi^2 2= 16.33, p< 0.001, \text{Kendall-W}= 0.54$)}. When asked to choose their preferred interaction, 26 participants chose Tangible, 3 chose Hand, and 1 chose both.
Subjects expressed their preferences for the prop using a 7-point Likert scale. They rated the ease of learning \textit{(M: 5.73 , SD: 1.36)}, responsiveness \textit{(M: 5.97 , SD: 0.99)} and modality input \textit{(M: 5.27 , SD: 1.53)}.

\subsection{Qualitative Feedback (Summary)}
Participants shared feedback on various aspects:

%\textit{Tangible Design:} Ergonomic concerns were common. \textit{P30} and \textit{P3} noted difficulty locating buttons during rotation, while \textit{P27} found the prop too heavy for extended use.
\textit{Tangible Design:} Ergonomic concerns were common. \textit{P\textsubscript{30,3}} noted difficulty locating buttons during rotation, while \textit{P\textsubscript{27}} found the prop too heavy for extended use.

%\textit{Scale Modal Switch:} Scaling activation was described as unintuitive by \textit{P21} and \textit{P1}, who suggested simplifying the input mechanism. \textit{P13} and \textit{P16} recommended clearer feedback through color and vibration.
\textit{Scale Modal Switch:} Scaling activation was described as unintuitive by \textit{P\textsubscript{21,1}}, who suggested simplifying the input mechanism. \textit{P\textsubscript{13,16}} recommended clearer feedback through color and vibration.

%\textit{Hand Interaction:} Gesture sensitivity and reliability were criticized by \textit{P12}, \textit{P17}, and \textit{P29}, who requested more responsive detection and shortcuts for simpler actions.
\textit{Hand Interaction:} Gesture sensitivity and reliability were criticized by \textit{P\textsubscript{12,17,29}}, who requested more responsive detection and shortcuts for simpler actions.

%\textit{Task-Specific Challenges:} Rotation was the most difficult action for \textit{P25} and \textit{P31} due to button placement. In combined tasks, \textit{P9} and \textit{P7} reported frustration with object selection and orientation visibility.
\textit{Task-Specific Challenges:} Rotation was the most difficult action for \textit{P\textsubscript{25,31}} due to button placement. In combined tasks, \textit{P\textsubscript{9,7}} reported frustration with object selection and orientation visibility.

%\textit{Overall Impressions:} Despite issues, tangible interaction was generally well-received. \textit{P9}, \textit{P19}, and \textit{P27} described it as natural, smooth, and easier to use than mid-air alternatives.
\textit{Overall Impressions:} Despite issues, tangible interaction was generally well-received. \textit{P\textsubscript{9,19,27}} described it as natural, smooth, and easier to use than mid-air alternatives.

\section{Discussion}
Our results should be understood in the context of the following limitations. Driven by our use case, we acknowledge that the tabletop condition and the constrained 4DoF task significantly influences the results of this study.
While the pilot study included mid-air tangible interaction, this was outside the scope of the main study, limiting our ability to disentangle the influence of the tabletop from the affordances of the TUI.
During the study, we noticed commonly observed variations in tracking accuracy of both hand and prop conditions. While prior work has evaluated tracking performance \cite{HMDAccuracy, MLtracking}, we found no studies specifically assessing hand and controller tracking accuracy or gesture responsiveness for the Magic Leap 2. Based on user feedback, the pinch hand gesture was perceived as less accurate. 
While prop tracking performed better, it may still have introduced minor inaccuracies.
Finally, the study focused on right-handed participants, and the results should be confirmed in a study with left-handed participants. 

\subsection{Rotation outperforms other scaling methods}
Our pilot study found that rotation-based scaling with TAN outperformed other methods, demonstrated by faster completion, lower errors, and positive user feedback.
This advantage likely stems from the lack of mid-air interaction, as table-constrained rotation is more stable and less fatiguing than lifting or hand-based methods, consistent with previous findings~\cite{fatigue}. The prop’s weight also favored rotation over vertical movement. Additionally, users may intuitively associate rotation with magnitude adjustment, as with everyday tools like knobs.
The cylindrical prop design further encouraged rotation through its natural affordance and grip, while hindering the LIFT method. However, this effect may vary with different shapes — e.g., a cube, might not favor rotation due to its different affordance and grip. Future studies should explore how prop shape and weight influence performance.

\subsection{Tangible outperforms hand in most tasks}
As expected, TAN outperformed HAND in completion time for all tasks except Only-Rot and Only-Scale, both involving prop rotation for TAN. In Only-Scale, HAND even surpassed TAN—a finding that contrasts with our pilot study on scaling methods (see Figure~\ref{fig:taskperformancescale}). This discrepancy could be due to the added modal-switch button, which was not present in the pilot. The design of mode switching and the button itself may have contributed to this disadvantage.
For the Only-Rot task, the select button's position may explain the lack of significant difference between TAN and HAND. When users rotated the prop through a large angle, the buttons end up in an odd position that is harder to reach. Moreover, the augmentation of the prop makes it difficult to visualize the current button positions. Both issues were noted in the qualitative feedback. A better design of the button layout may further improve the user experience and performance, leveraging button designs from other devices (see the survey \cite{buttons}). Rekimoto’s work on the Toolstone prop \cite{toolstone} reinforces this, showing how physical inputs can convey their state through touch, without relying on visual cues.

\subsection{Tangible is more precise}
%Reason why Combined task there were not sig.
%We found that TAN was more precise than HAND, except in the Combined task for rotation error. Interestingly, this pattern did not hold in the Move-And-Rotate task, despite it also involving multiple actions. A closer examination of rotation errors across tasks for each interactor reveals that TAN shows increasing error as task complexity rises—an outcome that may be expected—while HAND exhibits decreasing error. This may be due to the use of side handles in HAND, which participants were required to use in the Only-Rotate task. As tasks became more complex (Move-And-Rotate task), participants tended to rotate the object more while translating it. These findings suggest that the reduced precision in HAND is closely linked to the use of side handles.
% proposed shorter paragraph %
We found that TAN was more precise than HAND, except in the Combined task for rotation error. Interestingly, this pattern did not hold in the Move-And-Rotate task, despite also involving multiple actions. Examining rotation errors across tasks shows TAN's error increasing with task complexity—as expected—while HAND exhibits decreasing error. This may be due to HAND's side handles, which participants used in the Only-Rotate task. As tasks became more complex (Move-And-Rotate task), participants tended to rotate the object more while translating it. These findings suggest the reduced precision in HAND links to side handle use.

%Reflect to previous work
%Previous work has reported mixed findings when comparing hand and tangible interaction for precision. 
%Studies that dont align
%Some studies found contrasting results \cite{tansphere, inhandball, tanobjeval}, though it's worth noting that these comparisons involved different setups and tangible shapes. Only Bozgeyikli \cite{tanobjeval} directly compared tangible and hand interaction.
%Our knob-shaped design could have contributed to improved precision. However, this interpretation is debatable, as Voelker et al. \cite{knobtan} found no conclusive evidence that knobs enhance precision—though users did report a greater sense of control.
% proposed shorter paragraph %
Previous work has reported mixed findings when comparing hand and tangible interaction for precision. Some studies found contrasting results \cite{tansphere, inhandball, tanobjeval}, although they used different setups and tangible shapes. Only Bozgeyikli \cite{tanobjeval} directly compared tangible and hand interaction. Our knob-shaped design may have improved precision. Yet, Voelker \cite{knobtan} found no conclusive evidence that knobs resulted in less error—although error was calculated using overshoot metrics.
%Studies that align
%Our findings align with studies showing that cube- or block-shaped tangibles can outperform mid-air or flat interactions in terms of precision \cite{limp, tabletopdisplay}. As discussed earlier, table support played a key role in our design. The flat-bottomed tangible likely enhanced stability and control during manipulation. 
% proposed shorter paragraph %
Our findings align with studies showing cube/block tangibles outperform mid-air or flat interactions in precision \cite{limp, tabletopdisplay}. Table support played a key role in our design. The flat-bottomed tangible enhanced stability and control during use.

A complementary theory comes from research on embodied tangible sliders with constrained degrees of freedom (DoF), which showed improved precision \cite{embodiedaxes}. In our study, the 4DoF tasks, combined with the natural constraints of the table, likely reduced unnecessary movement and improved control. The increased precision observed with TAN may thus stem from biomechanical constraints introduced by both the tangible and the table, which limit the effective DoF of the user’s arm and hand—consistent with prior findings on motor control and grasping strategies \cite{posturalhand}.

Interestingly, although users were not required to keep the tangible on the table, many chose to do so. This suggests that both the tangible’s design and the presence of the table implicitly encouraged grounded interaction. In summary, we believe that both the table support and the ergonomic design of the tangible object contributed to the observed improvements in precision.

%Fatigue
Finally, fatigue effects should be considered. Although the average session lasted around 40 minutes (excluding training), tasks were repetitive but brief—insufficient to fully assess fatigue, which typically requires longer or a more demanding activity to manifest. However, over extended use, we hypothesize that performance differences would favor TAN. Unlike mid-air gestures, tangible objects can rest on a surface, reducing muscular strain and fatigue-related decline \cite{fatiguemidair}. This hypothesis is supported by task load ratings, which indicate early signs of increased physical demand in the HAND condition.
In conclusion, considering the performance and precision of TAN over HAND, we can partially accept \textbf{H1}.

\subsection{Task affects performance differences}
In this study, we analyzed manipulation tasks separately—first isolated tasks (Only-Move, Only-Rot and Only-Scale), then combined (Move-And-Rotate and Combined). The analyzed data show significant differences within these two groups.
For completion time, it was expected that the combined tasks would take more time than the isolated tasks, since they require a sequence of actions. We expected the completion time would not be the sum of time of the isolated tasks, since users could perform both tasks in parallel. Interestingly, for TAN there is no significant difference between Only-Rot and Move-And-Rotate.  Potentially,  TAN does enable parallel task performance better than HAND, since users need to rotate their wrist in order to rotate a manipulated object while moving the hand and maintain the hand parallel to the table. This goes in line with the aforementioned reduction of DoF that TAN offers.

%For the combined task, adding the scaling task to the Move-and-Rot significantly increased the task completion time, even further than their sums. Since the extra task cannot be performed in parallel, users needed to switch between tasks. This added to the extra complexity of the task dealing with other objects on the table, increased the mental demand of the tasks and in consequence its performance.  The increased cognitive load of multitasking compared to single task performance may be one explanation  \cite{multitasking}. Overall, these results are in line with previous work showing advantages of performing tasks in parallel, where multitasking performance is dependent on context like environmental demands, and the ability to switch between processing strategies \cite{parraleltask}.
% Proposed shorter version of this paragraph %
For the combined task, incorporating scaling into Move-and-Rot operations increased completion time beyond the sum of individual tasks. Since these operations couldn't be performed in parallel, participants switched between subtasks while navigating around table objects. This dual challenge increased mental demand and affected performance. As suggested in \cite{multitasking}, the cognitive load of multitasking likely explains these results. Our findings align with previous work showing that multitasking effectiveness depends on environmental context and strategy-switching abilities \cite{parraleltask}.

\subsection{Interaction method affects user patterns}
For the user interaction pattern, the performed analysis was mainly exploratory.
When analyzing the spatial position of TAN and HAND, it is clear that TAN and its supportive table influenced how the user interacted with the virtual object. The movements were smoother and along the ideal path. This pattern is likely  due to the decreased number of DoF the TAN condition imposes. 
%*** WHY? ***
In contrast, HAND tended to result in longer trajectories with greater deviations from the ideal path and more corrective movements. The misalignment and corrective pattern might increase with longer tasks, as it will likely be affected by user fatigue. This can be complemented by the results of Cheng \cite{tabletopErgo}, where posture and ergonomics were analyzed. The study found that decreased movement of tangibles may lead to a more controlled and stabilized task performance.
%** delve deeper in that article here***

Analysis of the task sequence plot for the combined task reveals visual evidence that TAN and HAND exhibit distinct patterns consistent across users. HAND presented more interactions than tangible. This is also demonstrated by the overshoot count metric. We believe that to achieve the task with good precision, subjects needed more corrective steps. One reason could be the selection logic that is different for both interactions. For HAND, users needed to keep the pinch gesture to perform the task; for TAN the engaging button behavior frees up the hand for the manipulation.

When analyzing the sequence of active tasks, both TAN and HAND seem to have mainly Move-And-Rotate and Scale active tasks. Only-Move or Only-Rot are rarely performed, which is in line with the  parallel task performance  both techniques enable. For Only-Scale, HAND presented a fair amount of Scaling activity, and took longer than in TAN. This is in contrast to the completion times for Only-Scale, but can answer why the combined task take longer. Scaling pattern is quite different between HAND and TAN. If we look closer, scaling occupies less time in the complete task chain compared to move and rotate. Active scaling (not considering activating the mode), is quite fast and presumably precise, likely explaining this finding.  In conclusion, the data support \textbf{H2}.

In summary, our study revealed several key findings. 
%HAND FINDINGS
Hand interactions demonstrated efficiency in isolated scaling tasks, which were completed significantly faster. However, in more complex combined tasks, users made multiple adjustments, resulting in increased time spent. 
%TAN FINDINGS
For tangible interactions, we found that rotation-based scaling outperformed other methods, likely due to the knob-shaped affordances. Yet, incorporating scaling into the combined task consistently degraded performance across all metrics. This decline was attributed to the introduction of a modal switch, where the same physical rotation was used to control both object rotation and scaling. Participants found this added complexity to be cognitively demanding. 
Finally, we demonstrated that tangible interactions were generally more precise than hand-based ones. This increased precision can be attributed to the physical design of the prop—particularly its flat-bottomed shape, which leveraged the stability of the table surface—and the reduced degrees of freedom required for the task.

\section{Conclusion}
This study offers insights into how users manipulate virtual 3D objects along 4DoF on real surfaces in mixed reality (MR) using hand and tangible interactions. By comparing isolated and combined tasks, we found that hand interactions enabled faster isolated scaling, but combined tasks required more time due to repeated adjustments. In contrast, tangible interactions suffered when scaling was added, likely due to the complexity introduced by a modal switch for knob-based scaling and rotation. Despite this, tangible interactions led to faster completion times in full 4DoF tasks and consistently higher precision. These findings highlight the importance of analyzing both isolated and complex tasks to understand trade-offs between interaction modalities. Tangible interfaces show strong potential for MR applications requiring fine-grained manipulation, such as plant and room planning. Future work should aim to reduce interaction complexity to further improve performance.

%This study opens several avenues for future research. First, extending the interaction space from the current 4DoF (2D translation, 1D rotation, 1D scaling) to full 9DoF manipulation would enable more complex object interactions, particularly in 3D environments. Additionally, exploring bi-manual interaction techniques could enhance efficiency and expressiveness, especially for tasks that benefit from simultaneous manipulation and control. Another promising avenue is the use of multiple tangible objects, where each tangible could serve a distinct function or mode. For instance, inspired by Rekimoto's Toolstone metaphor \cite{toolstone}, one tangible could act as a mode switcher, enabling users to fluidly transition between interaction types. These extensions would not only broaden the applicability of the system but also provide deeper insights into the trade-offs between tangible and hand-based interaction paradigms.
% proposed shorter paragraph %
This study opens several avenues for future research. First, extending the interaction space from 4DoF (2D translation, 1D rotation, 1D scaling) to full 9DoF manipulation would enable more complex object interactions, particularly in 3D environments. Additionally, exploring bi-manual interaction techniques could enhance efficiency and expressiveness, especially for tasks that benefit from simultaneous manipulation and control. Another promising avenue is the use of multiple tangible objects, where each could serve a distinct function or mode. For instance, inspired by Rekimoto's Toolstone metaphor \cite{toolstone}, one tangible could act as a mode switcher, enabling users to fluidly transition between interaction types. These extensions would broaden the system’s applicability and provide deeper insights into tangible versus hand-based interactions.

\bibliographystyle{ACM-Reference-Format}
\bibliography{tangible-bib}

\end{document}